\documentclass[10pt,amsmath,amssymb,aps,floatfix,prl,tightenlines,twocolumn]{revtex4-2}

\usepackage{import}
\usepackage{natbib}
\usepackage[pdftex]{graphicx} 
\usepackage{physics}
\usepackage{multirow}
\usepackage{xcolor}
\usepackage{dcolumn}
\usepackage{bm}
\usepackage[normalem]{ulem} 
\usepackage{float}

\begin{document}

\title{Quantum mechanics can find a needle in a haystack every time}

\author{F. Mohit$^{1,2}$, J. Guanzon$^{1,2}$, J. McKinlay$^{1,2}$, T.~J. Weinhold$^{1,2}$, C.~R.~Myers$^{3,4}$, M. P. Almeida$^{1,2}$, M. Rambach$^{1,2}$, and A.~G. White$^{1,2}$}
\affiliation{$^{1}$Australian Research Council Centre of Excellence for Engineered Quantum Systems\\
$^{2}$\mbox{School of Mathematics and Physics, University of Queensland, Brisbane, Australia}\\
$^{3}$\mbox{School of Physics, Mathematics and Computing, The University of Western Australia, Perth, Australia}\\
$^{4}$\mbox{Pawsey Supercomputing Centre, Perth, Australia}}
\date{\today}

\begin{abstract}   
\noindent
Grover's algorithm is one of the pioneering demonstrations of the advantages of quantum computing over its classical counterpart, providing---at most---a quadratic speed-up over the classical solution for unstructured database search. The original formulation of Grover's algorithm is non-deterministic, finding the answer with a probability that varies with the size of the search space and the number of marked elements. A recent reformulation introduced a \emph{deterministic} form of Grover's algorithm that---in principle---finds the answer with certainty. Here we realise the deterministic Grover's algorithm on a programmable photonic integrated circuit, finding that it not only outperforms the original Grover's algorithm as predicted, but is also markedly more robust against technological imperfections. We explore databases of 4 to 10 elements, with every choice of a single marked element, achieving an average success probability of $99.77 \pm 0.05\%$.

\end{abstract}

\maketitle
Quantum computers offer the potential to tackle problems---leveraging quantum entanglement and superposition---that would be impractical or highly inefficient using classical computation methods~\cite{Nielsen_Chuang_2010, Preskill2012}.  A key example of this is quantum search algorithms, pioneered by Grover~\cite{OriginalGrover}. In a variety of fields a regular task is to search an unstructured database for marked elements, examples include: searching for a specific gene sequence in a genomic database; fuzz testing to find inputs that make a computer program crash; or finding a password that matches a specific hash function. By efficiently searching such databases via amplitude amplification, Grover's algorithm provides a quadratic acceleration over the best-known classical search algorithm for large database sizes: a one-day classical data search with a query every second would only take around five minutes with Grover's quantum algorithm~\cite{Zalka1999, PhysRevA.102.042609}. This polynomial speed-up is optimal~\cite{Zalka1999}.

The original formulation of the algorithm is non-deterministic:~the probability of success of finding a marked element varies with the size of the search space and the number of marked elements, and can be as low as 87\%. This non-deterministic Grover's algorithm has been realised in an array of physical architectures~\cite{KwiatJMO, PhysRevLett.88.137901, DAS20038, PhysRevA.72.050306, HostenNature2006, Hijmans:07, DiCarloNature, Perez-Garcia20092018, PhysRevA.102.042609, Heurtel2023percevalsoftware} over the last 25 years. Various deterministic formulations of the algorithm have been found which require access to the oracle \cite{long_grover_2001, brassard_quantum_2002, hoyer_arbitrary_2000}. This limits their usage as the oracle may not always be accessible. A deterministic version of Grover's algorithm with a \textit{restricted oracle} was proposed in 2022~\cite{roy_deterministic_2022}: the first and so far only deterministic formulation not requiring access to the oracle, and works as long as the ratio between the database size, $N$, and the number of marked elements $M$ is $N/M \geq 4$. It has been experimentally demonstrated in bulk optics \cite{he_experimental_2023} for $N/M {=} 4$ (for which the original algorithm is deterministic) and $N/M {=} 8$; and in integrated photonics \cite{li_experimental_2023} for $N/M {=} 6$.

\begin{figure}[!b]
\centering
\vspace{-6 mm}
\includegraphics[width=0.55\columnwidth]{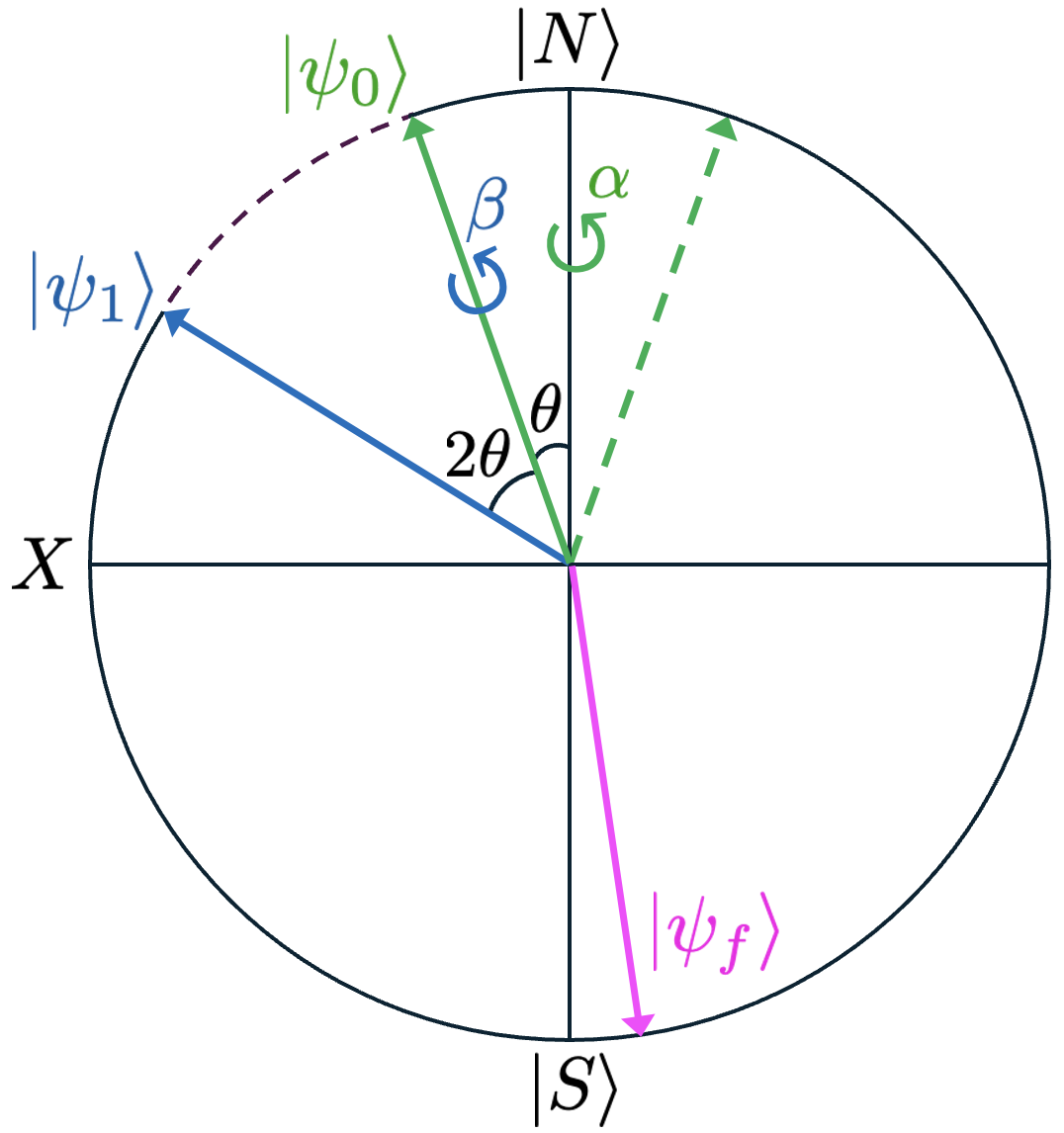}
\vspace{-3 mm}
\caption{\emph{Original Grover's algorithm}. The initial state of the system (solid green arrow, $|\psi_0\rangle$), is rotated to the final state (pink arrow, $|\psi_f\rangle$) via iterations involving a phase flip operation on all database elements (dashed green arrow) followed by a diffusion operator or reflection about the initial state (also known as the amplitude amplification step), ending the first iteration at the solid blue arrow, $|\psi_1\rangle$. Each iteration rotates the state of the system towards the final state by $2\theta$. The north and south poles, $\ket{N}$ and $\ket{S}$, correspond to superposition states of all unmarked and marked elements, respectively. Figure adapted from \cite{roy_deterministic_2022}.} 
\label{fig:Bloch1}
\vspace{-2mm}
\end{figure}

In this work we present a demonstration of the Roy et~al. deterministic Grover's algorithm~\cite{roy_deterministic_2022} on two integrated photonic processors, respectively 20 and 12 modes, showing improved success probabilities for all $N/M{>}4$ on both systems. We implemented a sequential optimisation technique to improve the 20-mode processor performance to a local optimum and used a clear-box  optimisation technique---presented in \cite{Fyrillas2024}---to optimise the 12-mode processor, achieving near-optimal performance in the latter case. In all cases we found increased robustness of the deterministic algorithm compared to the original algorithm with respect to experimental imperfections.

To understand how the deterministic version of Grover's algorithm works, we first consider the original algorithm: given a database of $N$ elements, the system is initialised by creating an equal superposition of all elements, e.g. creating the state $\ket{\psi_0} = |1,1,1,1,...\rangle$. In the 2-dimensional representation of the system in Fig.~\ref{fig:Bloch1}, $\ket{\psi_0}$ corresponds to the solid green arrow. Here, the north ($\ket{N}$) and south ($\ket{S}$) poles represent superpositions of all \textit{unmarked} and \textit{marked} elements of the database, respectively.

\begin{figure}[t]
\centering
\includegraphics[width=\columnwidth]{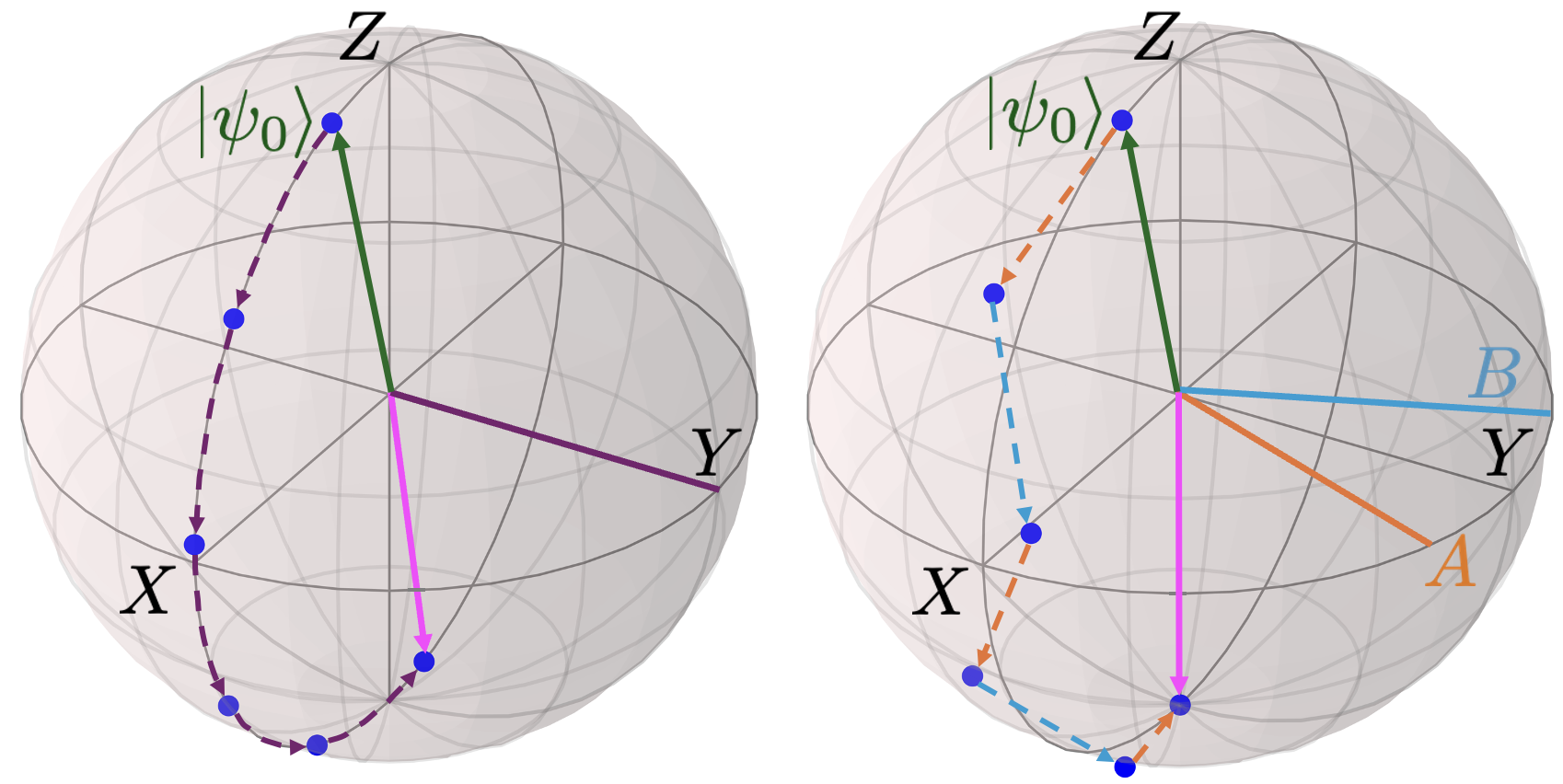}
\vspace{-6mm}
\caption{Bloch sphere representations of: 
\emph{left}: \emph{Original Grover's}: the diffusion operator is applied about a fixed axis (Y) in every iteration. This results in the final state not aligning perfectly with the target state $\ket{S}$, and the misalignment depending on the ratio of marked items to total inputs.
\emph{right}: \emph{Deterministic Grover's}: two axes (A and B) are calculated, about which the diffusion operator is applied alternatively in each iteration. The final state is constrained to the desired state $\ket{S}$. Figure adapted from \cite{roy_deterministic_2022}.}

\label{fig:Bloch2}
\vspace{-6mm}
\end{figure}

The algorithm's goal is to rotate the initial state of the system until it is aligned with the desired final state, $\ket{S}$. This is done iteratively through two steps, the oracle query and amplitude amplification. In the \emph{oracle query} step, a phase flip operator, known as the oracle, is applied to the entire system---following initialisation---which negates the phase of marked element(s) and is identity for all other elements, resulting in a rotation around the vertical axis by an angle $\alpha{=}\theta$, to the state represented by the dashed green arrow in Fig.~\ref{fig:Bloch1}. \emph{Amplitude amplification}, applies a second rotation, $\beta{=}2\theta$---known as the diffusion operator---around the initial state of the system, $\ket{\psi_0}$. This brings the state of the system to the solid blue arrow in Fig.~\ref{fig:Bloch1}, closer to the desired final state, $\ket{S}$. This process is then repeated until the algorithm converges at the solid pink arrow. 

This original algorithm is probabilistic, finding the marked element with a success probability that depends non-trivially on the size of the database and number of marked elements. In each iteration, the state of the system moves closer to the desired final state by $2\theta$, as shown in  Fig.~\ref{fig:Bloch1}. Therefore, the total number of iterations is $(\pi{-}\theta)/2\pi$, which is usually non-integer. As the number of iterations is an integer, this ideal value is rounded down to the nearest whole number, resulting in the final state not being perfectly aligning with the south pole in all cases where the ideal number of iterations was a non-interger number. For example, with a single marked element, the success probability is 100\% for $N{=}4$, drops to 87\% for $N{=}7$, is near 100\% for $N{=}9$, and is below 94\% for $N{=}15$.

In the deterministic version of Grover's algorithm~\cite{roy_deterministic_2022}, the marked element(s) are found with 100\% probability of success for \emph{any} database size as long as $N/M {\geq} 4$. The difference between the algorithms arises in the diffusion operator, which in the original algorithm is applied about a fixed axis in every iteration, e.g. the Y-axis as shown on the left side of Fig.~\ref{fig:Bloch2}. It can be shown that by constraining the final state of the system to align with the south pole, two alternative rotation axes can be calculated for the diffusion operator, ensuring the deterministic success of the algorithm every time. In the deterministic algorithm, two diffusion operators are applied alternatively about axes A and B, as shown on the right side of Fig.~\ref{fig:Bloch2}. This results in, at most, $\kappa {+} 1$ iterations, where $\kappa$ is the number of iterations in the original Grover's algorithm.

To directly compare performance of the original and deterministic Grover's algorithms , we implement them both on the same experimental architecture: a programmable photonic integrated platform using single photons. We use two platforms: an in-house platform and a cloud platform from Quandela. The processors in both platforms are programmable SiN quantum photonic circuits  manufactured by Quix Quantum (QUIX Quantum)~\cite{Taballione_2021,Taballione2023} with 20 and 12 spatial modes for the in-house and cloud systems respectively.

\begin{figure*}[t]
\centering
\includegraphics[width=1.7\columnwidth]{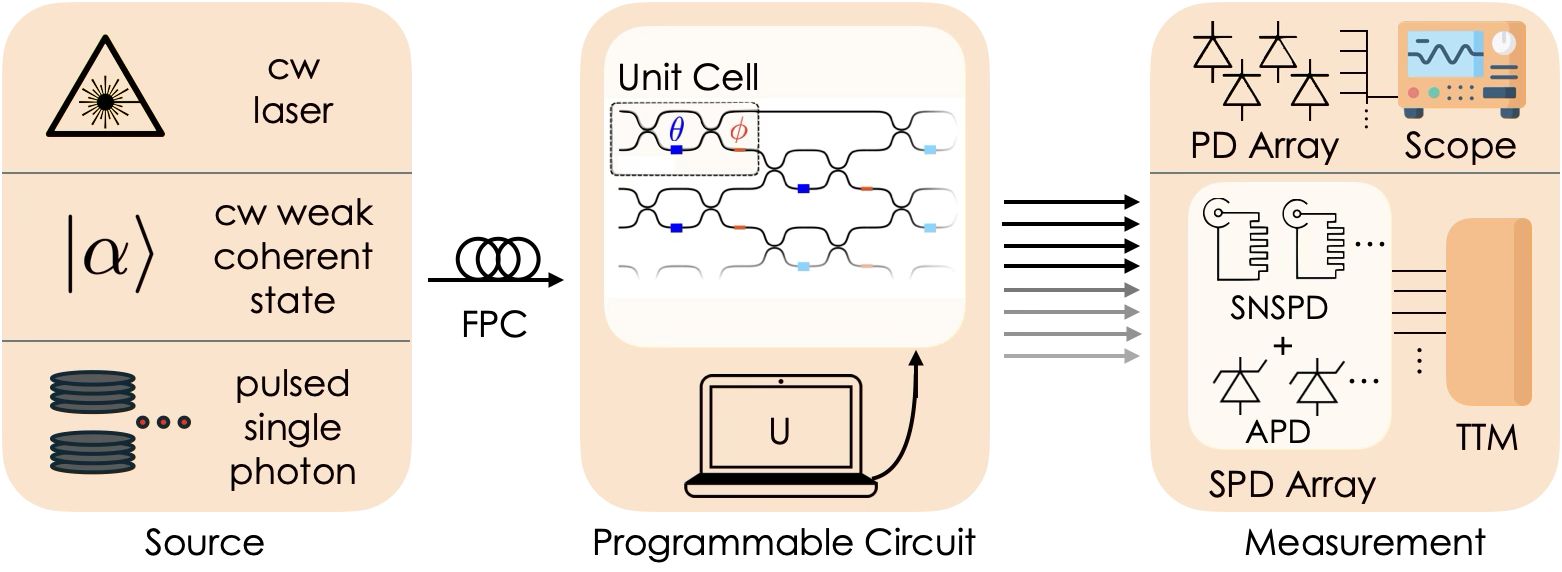}
\vspace{-3mm}
\caption{Schematic for in-house experimental platform. Three sources, all at $\sim$925 nm, are variously used: for calibration, a continuous wave laser, CW, and a weak coherent state, $|\alpha \rangle$, achieved using a fibre attenuator on the CW laser; for final data, a quantum-dot single-photon source. Any of these sources are connected to the first channel (spatial mode) of the programmable circuit after passing through a fibre polarisation controller (FPC) to match the polarisation orientation accepted by the circuit. The unitary transformations in both the deterministic and original algorithms, $U$, are set via software. The circuit consists of a mesh of unit cells containing two fixed 50:50 beam splitters and two phase shifters. These control the splitting ratio (blue) and relative phase (red) between two neighbouring modes via angles $\theta$ and $\phi$, respectively. The output of relevant modes 1--8 are measured using the appropriate detectors for each source: for the CW laser an array of $N$ photodiodes (PDs); for both the weak coherent source and the single photon source a single photon detector (SPD) array, consisting of up to $N{=}4$ superconducting nanowire single-photon detectors (SNSPDs) and, for $N{>}4$, $N{-}4$ avalanche photodiodes (APDs). An oscilloscope and time-tagger module (TTM) are used for signal read-out for the photodiodes and single-photon detectors, respectively. \vspace{-5mm}}
\label{fig:exp}
\vspace{0mm}
\end{figure*}

Figure \ref{fig:exp} outlines our experiment performed with our in-house platform. We have a choice of light sources, either: a) a continuous-wave laser at 925~nm (Toptica CTL)---used either as a bright source for circuit characterisation or as a source of weak coherent states, $\alpha \ll 1$; or b) a solid-state cavity quantum-dot pulsed source producing sub-ns single-photons at 925.1~nm at a repetition rate of 80~MHz (Quandela e-Delight NR)~\cite{quandela-source}. Either of these light sources is input into one channel of the processor. Fibre polarisation controllers (FPC) were used at input to optimise light throughput, since the processor is polarisation-sensitive.

The processor implements arbitrary linear transformations on input states of light using 380 phase shifters in a mesh of Mach-Zehnder interferometers, shown schematically in the central panel of Fig.~\ref{fig:exp}. Each unit cell of the programmable circuit consists of a programmable beamsplitter---a Mach-Zehnder interferometer with embedded phase-shifter, $\theta_i$---followed by a further programmable phase shifter, $\phi_{i}$. The phase shifters are heaters patterned on the chip: varying the current through the heater varies the refractive index of the SiN waveguide, controllably inducing a phase shift on that waveguide. Overall chip temperature is maintained at $40^\circ\mathrm{C}$ by a Peltier cooling system. The SiN processor is designed for optimal performance at 940~nm, with the losses from each input mode to the respective output mode ranging between 3--7~dB, with a median loss of $4.2$~dB. At our operational wavelength of ${\sim}925$~nm, the loss range is similar (see Supplemental Material for further details on the loss~\cite{Supp}).

The output light is measured using appropriate detectors: photodiodes for the bright laser and a combination of superconducting nanowire (SNSPD) and avalanche photodiodes (APD) for either the weak coherent or single-photon inputs. Each output of the experiment is efficiency corrected based on its fibre coupling, $\eta_{\mathrm{c}}$, and detector efficiency, $\eta_{\mathrm{d}}$, the latter differing significantly between the SNSPDs and APDs. We first send attenuated CW light through one reference input channel (Ch.~1) and re-route it to different outputs. We then calculate correction factors, $\eta_{\mathrm{tot}}^{-1} \, {=} \, 1/(\eta_{\mathrm{f}} \, \eta_{\mathrm{d}}) \, {\geq} \, 1$, to account for the total efficiency of individual output channels. See supplementary material for further details on efficiencies and correction factors~\cite{Supp}.

We implement Grover's algorithm by inputting the light into one channel of the processor and then superposing that light across $N$ channels of the processor with a unitary $U_{N}$. Grover's algorithm then consists of marking one of those channels by adding a $\pi$ phase shift---this is the Oracle---and then performing amplitude amplification $U_{A}$. This is repeated at most $\kappa{+}1$ times. We combine all of these operations into a single unitary, $U_{\mathrm{G}}$ which we program into the circuit. 

In principle, any given unitary, $U$, can be realised by a Mach-Zehnder network~\cite{Reck1997}, with the Clements et al. decomposition~\cite{clements_optimal_2016} used here to determine the values of the phase shifters, $\theta_{i}$ and $\phi_{i}$, for each unit cell $i$. In practice, however, due to the complex structure of the chip---the interplay between the heating wire architecture and the layout of the photonic waveguides---the processor is subject to significant cross-talk effects. For example, the phase shift at heater $\theta_{i}$, is modified depending on the heater values of unassociated phase shifts, $\theta_{j \neq i}$ and $\phi_{1...\mathcal{N}}$, where $2\mathcal{N}$ are the total number of heater elements. This is recognised by the manufacturer, which supplies a Python package that carries out the ideal decomposition to set phase shifter values and then varies some of the values based on the manufacturer's knowledge of cross-talk.

In addition, the fixed beamsplitters---in the parlance of photonic integrated circuits these are known as directional couplers---although nominally 50\%, vary by up to ${\sim}4.5$\% from this, moving the actual implementation even further from the ideal. In general, if the beamsplitters are imbalanced, then for one of the interferometer inputs there is no phase setting, $\theta$, that will fully route all the input light to any given output. Given these caveats, there will be an unavoidable difference between the measured and ideal outputs for any non-trivial unitary.

\begin{figure}[!t]
\centering
\includegraphics[width=0.91 \columnwidth]{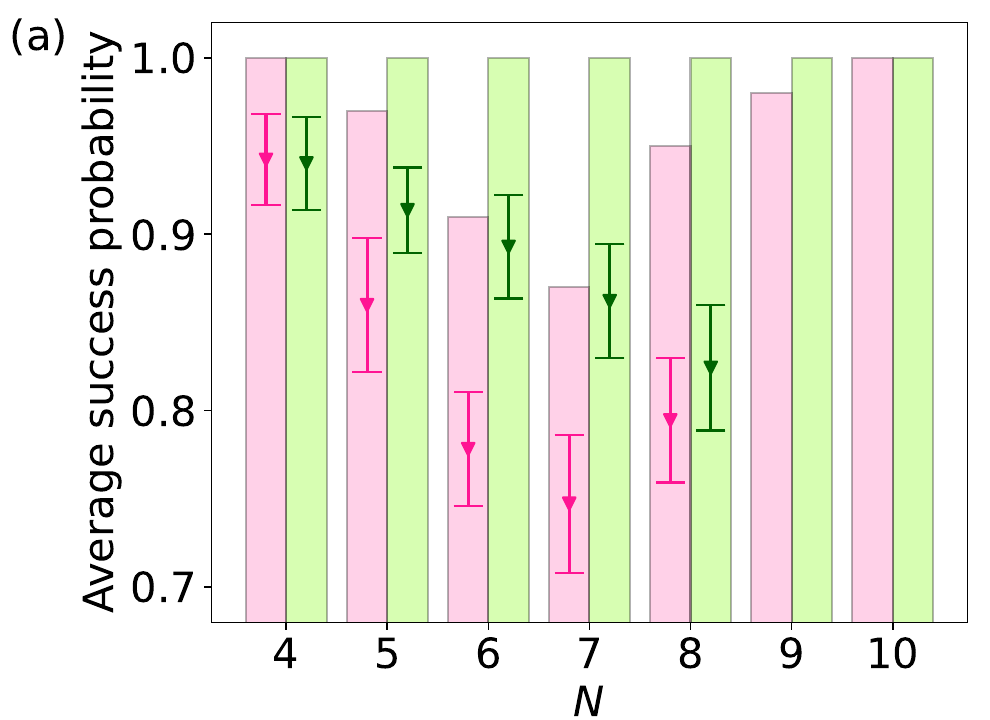}
\includegraphics[width=0.91 \columnwidth]{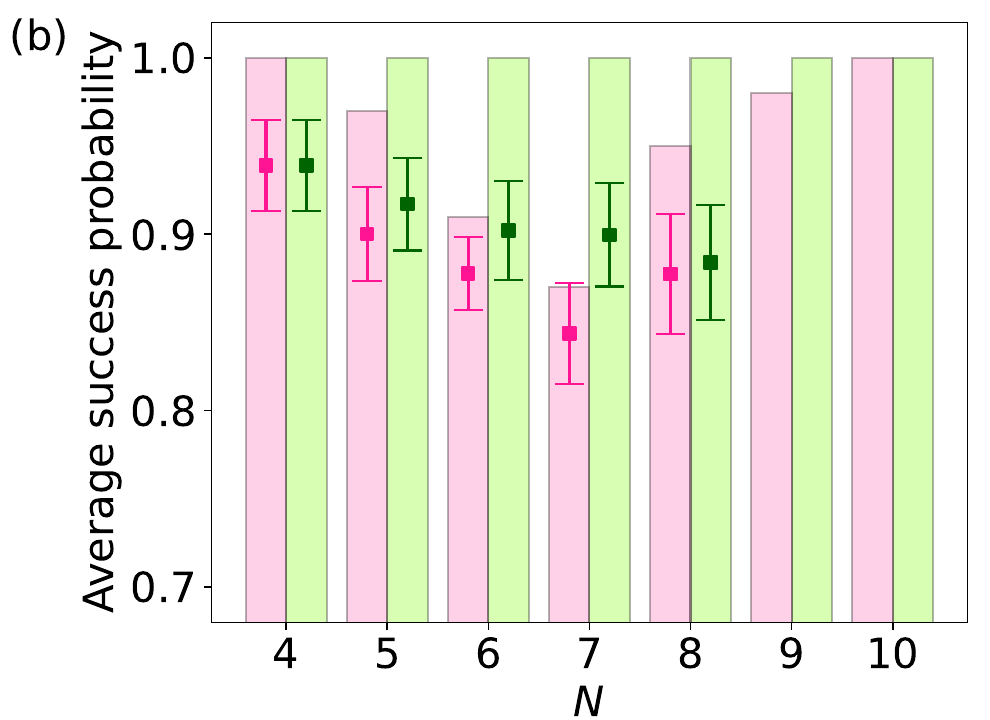}
\includegraphics[width=0.91 \columnwidth]{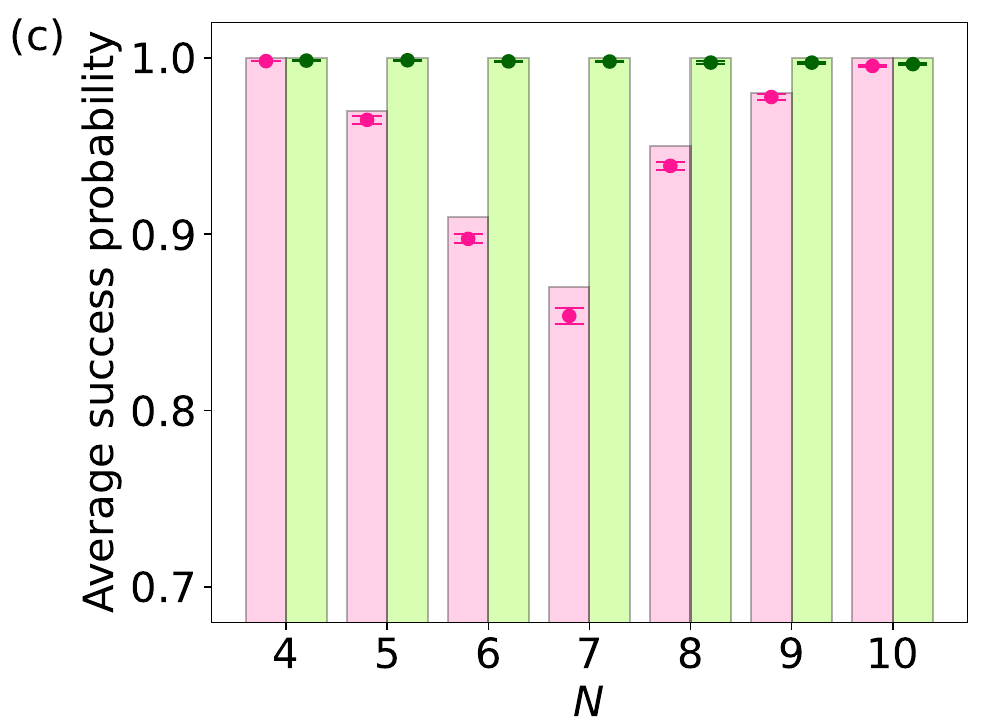}
\vspace{-4mm}
\caption{Assessment of original vs deterministic Grover's algorithm. The coloured bars are the ideal success probabilities for the original (pink) and deterministic (green) Grover's algorithms with one marked element in databases with $N{=}4$ to $N{=}10$ elements. The data points are the average experimental probabilities for the original (dark pink) and deterministic (dark green) Grover's algorithms. These are obtained by respectively marking each element of the database, 1 through $N$, and then measuring the subsequent success probability; the error bars are the standard deviation in the average. (a) using an unoptimised 20-mode processor and classical light sent into channel 1 and deteced via avalanche photodiodes. (b) using sequentially-optimised 20-mode processor and single photons that are sent into channel 1 of the processor and detected using $N{=}4$ nanowire + $(N{-}4)$ avalanche detectors. (c) using clear-box optimised 12-mode processor and single photons, again sent into channel 1. \vspace{-7mm}}
\label{fig:MeasProbs}
\end{figure}

Following this, we found that the manufacturer's compensation still led to an imperfect implementation of the algorithms and that additional compensation was necessary to gain optimal outcomes. To get an initial assessment of chip performance, we inputted the continuous-wave laser and used photodiodes to measure the output intensities, obtaining the results shown in Fig.~\ref{fig:MeasProbs}a. When using narrow linewidth light sources there are etalon effects at fibre connections that lead to significant phase drift and intensity changes: to avoid this we reduced the effective coherence length of the laser by sinusoidally dithering its wavelength. 

We can see that both the original and deterministic algorithms rapidly diverge from the ideal results as the database size, $N$, is increased. Closer observation shows that the deterministic Grover's suffers less degradation than the original Grover's, e.g.~at $N{=}6$, the original Grover's is 84\% of the theoretical value c.f.~89\% for the deterministic Grover's. 

Given these promising initial results, we chose to develop a sequentual optimisation routine to mitigate cross-talk and beampslitter effects. We evaluate the difference between expected and measured output intensity distributions using the total variation distance (TVD),
\begin{equation}
    \text{TVD} = \frac{1}{2}\sum|p_{i} - q_{i}|,
\end{equation}
where $p_{i}$ is the theoretically-expected, and $q_{i}$ is the experimentally-measured, values of the normalised output intensity distributions. A TVD of 0(1) corresponds to zero(maximal) difference between theory and experiment. The optimisation process minimises the TVD through individual tuning of the heaters. 

\begin{figure}[!t]
\centering
\includegraphics[width=\columnwidth]{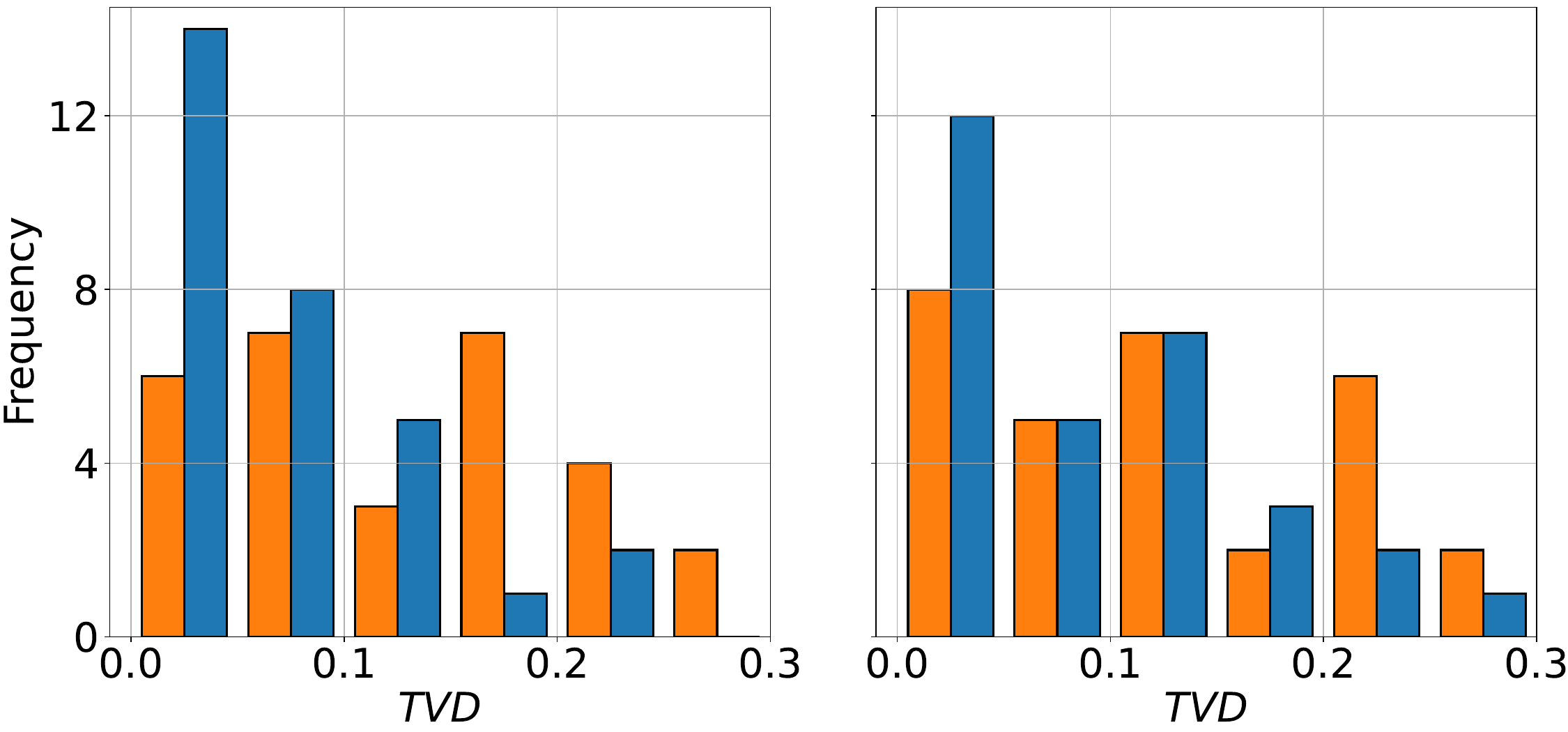}
\vspace{-6mm}
\caption{Frequency of total variation distance (TVD) for all marked elements and all database sizes measured using the in-house system, 30 measurements in total. \emph{left}: original, and \emph{right}: deterministic, algorithms. Orange bars indicate values for unoptimised circuits, blue bars indicate values for optimised circuits. Circuit optimisation leads to notably lower TVD values, as desired.
}
\label{fig:TVDhist}
\vspace{-6mm}
\end{figure}

The first step in the optimisation is to find and compile a list of the heaters engaged in implementing the unitary transformations for each algorithm. Out of 380 heating elements, our unitary sizes of $N{=}4$ up to $N{=}8$ use at least 12 and up to 56 heaters. Next, we sequentially target each phase-shifter and vary its phase by a small increment. If this change results in an improvement in the TVD, then the phase of the target heater is updated. We repeat the process for the same heater until no further improvement is achieved, within a pre-specified accuracy to avoid the noise floor. Finally, all heaters used to implement the transformation are consecutively optimised in this way. In the majority of cases, there was little to no change in the phase of the heaters, and for those that were notably varied, the majority of changes were within a small range of the original value. See supplementary material for further details on the optimisation routine~\cite{Supp}. 

The optimised algorithm was then assessed with both classical and quantum light, using:~1)~the continuous-wave laser measured using conventional photodiodes, as above;~2)~the continuous-wave laser attenuated down to single-photon level, $\alpha {\ll} 1$, and measured using single-photon detectors; and 3) a quantum-dot single-photon source measured using single-photon detectors. All were consistent with each other, so from here on, we consider only the single-photon source results.

Fig.~\ref{fig:TVDhist} captures the improvement in circuit performance due to optimisation. The left and right histograms are for the original and deterministic algorithms, respectively. The orange bars are for the unoptimised circuit, showing the TVD frequency for all marked elements, 1 through $N$, for all database sizes, $N{=}4$ through $N{=}8$, a total of 30 measurements. The blue bars are the frequencies after optimisation, and show a notable reduction in high TVD values and a marked increase in low TVD values, as desired.

Table~\ref{tab:TVD} is a finer-grained examination of the improvements, listing the median TVD for each database size $N$, before and after optimisation. When unoptimised, the original algorithm had a median TVD ranging from 0.05--0.18; when optimised, this reduced significantly to 0.04--0.07. We see similar improvements for the deterministic algorithm, with the median TVD of the optimised circuit ranging from 0.05--0.21; when optimised, this again reduces significantly to 0.05--0.08. 

\begin{table}[b]
\begin{tabular}{|l|l|l|l|l|l|l|}
\hline
\textbf{Database size (N)}           &                & 4               & 5               & 6               & 7               & 8               \\ \hline \hline
\textbf{\ Median TVD}      &         &        &                 &                 &                 &                 \\ \hline 
\multirow{2}{3em}{Original} & unoptimised & 0.05 & 0.10             & 0.15            & 0.16            & 0.18      \\ 
& optimised & 0.05 & 0.07            & 0.04            & 0.05            & 0.05  \\ 
\hline
\multirow{2}{3em}{Deterministic} & unoptimised & 0.05            & 0.09            & 0.10             & 0.13            & 0.21      \\ 
& optimised & 0.05            & 0.06            & 0.08            & 0.06            & 0.08  \\ 
\hline           
\end{tabular}%
\vspace{-1mm}
\caption{Effect of optimisation on median total variation distance (TVD) for both original and deterministic algorithms for each database size.
}
\label{tab:TVD}
\end{table}

In Fig.~\ref{fig:MeasProbs}b we can see the optimised original algorithm average success probabilities are significantly closer to the theoretical amplitudes, starting at maximum intensity for $N{=}4$ and gradually decreasing to a minimum at $N{=}7$ before increasing again. The optimised deterministic average success probabilities are also significantly closer to the theoretical uniform distribution across $N$ while still outperforming the original algorithm. (Table~\ref{tab:successprobs} in the Supplemental Material~\cite{Supp} details the numerical values for the theoretical success probabilities, original, and deterministic average success probabilities from this Figure). This optimisation approach, while quick, straightforward and effective, does not leverage any knowledge about the underlying physical system of the processor, giving rise to average success probabilities that are still non-optimal despite significant improvement. 

In the final part of the experiment, we use a clear-box optimisation approach from \cite{Fyrillas2024} on a 12-mode cloud processor to reach a global optimum for these values. The optimisation technique in this part models the relationship between phase values to the voltage values of individual heaters while compensating for possible sources of error arising from cross-talk, imbalanced beam-splitters and varying insertion loss across spatial modes of the circuit. The model is given by:
\begin{equation}
    \Vec{\phi} {=} C_2 \cdot \Vec{V}^{\odot 2} + \Vec{c_0},
    \label{eqn:optim}
\end{equation}
where $\Vec{\phi}$ are the phase values, $\Vec{V}$ are the voltage values, $C_2$ is a matrix representing the effects of cross-talk and $\Vec{c_0}$ are a set of passive phases that are used to correct for varying insertion loss across different modes. An Adam optimiser is used to train the model on a set of randomly generated voltage values via stochastic gradient descent \cite{Fyrillas2024}. This process finds the relevant parameters in Equation \ref{eqn:optim} which can then be used to predict the required voltage values given a set of desired phases to implement on the interferometer. For further details on this optimisation process, see \cite{Fyrillas2024}. We implemented both the original and deterministic Grover's algorithms on this system for database sizes $N{=}$\,4--10. 

The average success probabilities of the fully optimised system in Fig.~\ref{fig:MeasProbs}c are far closer to theory than the previous unoptimised and partially optimised results. Averaged over all database sizes, $N{=}$4--10, and over all combinations of marked elements, the average success probability is $99.77 \pm 0.05$\%. Note that the deterministic algorithm consistently outperforms the original algorithm, with a much smaller error for each individual database size: reinforcing the robustness of the deterministic algorithm when compared to the original. This performance exceeds that of all previous demonstrations of Grover's algorithm \cite{citekey,mandviwalla_implementing_2018, hlembotskyi_efficient_2020, gwinner_benchmarking_2021, zhang_implementation_2021, j_quantum_2022,  zhang_quantum_2022, thorvaldson_grovers_2025}.

Two trends are evident. Firstly, in terms of success probability, the deterministic algorithm always outperforms the original algorithm for all three implementations: unoptimised, sequentially-otpimised, and clear-box  optimised. Secondly, the deterministic algorithm tends to be more robust to circuit imperfections, in a majority of cases being much closer to the expected ideal values than the non-deterministic version. The former trend is as expected from theory, the latter trend is a reason to implement the deterministic Grover's algorithm even if higher success probability is not a concern or desiderata. 

Open avenues of research include exploring whether these advantages are maintained when Grover's algorithm is part of larger photonic computation, which experimentally will mean the search photon is entangled to other computational photons, or if these advantages also accrue when Grover's algorithm is realised with other physical encodings, such as superconducting qubits \cite{mandviwalla_implementing_2018, zhang_implementation_2021, j_quantum_2022, gwinner_benchmarking_2021, citekey}, trapped ions \cite{hlembotskyi_efficient_2020, zhang_quantum_2022}, or spin qubits \cite{thorvaldson_grovers_2025}. 

\noindent \emph{Acknowledgements}. We thank Quix Quantum for the loan of the 20-mode quantum photonic processor, and Jonas Philipps, Michiel de Goede, and Nikhil Sen for helpful discussions. This research was supported by the Australian Research Council Centre of Excellence for Engineered Quantum Systems (EQUS, CE170100009). AGW is the recipient of an Australian Research Council Laureate Fellowship (FL210100045) funded by the Australian Government.

\vspace{-1mm}
%

\newpage
\section{Supplementary Material:\\
Quantum mechanics can find a needle in a haystack every time}
This supplementary material is with reference to our in-house 20-mode photonic processor and the experimental set-up associated with it.

\vspace{3mm}
\noindent \textbf{\emph{Losses in programmable circuit}}
\vspace{2mm}

\begin{figure}[b!]
\centering
\includegraphics[width=\columnwidth]{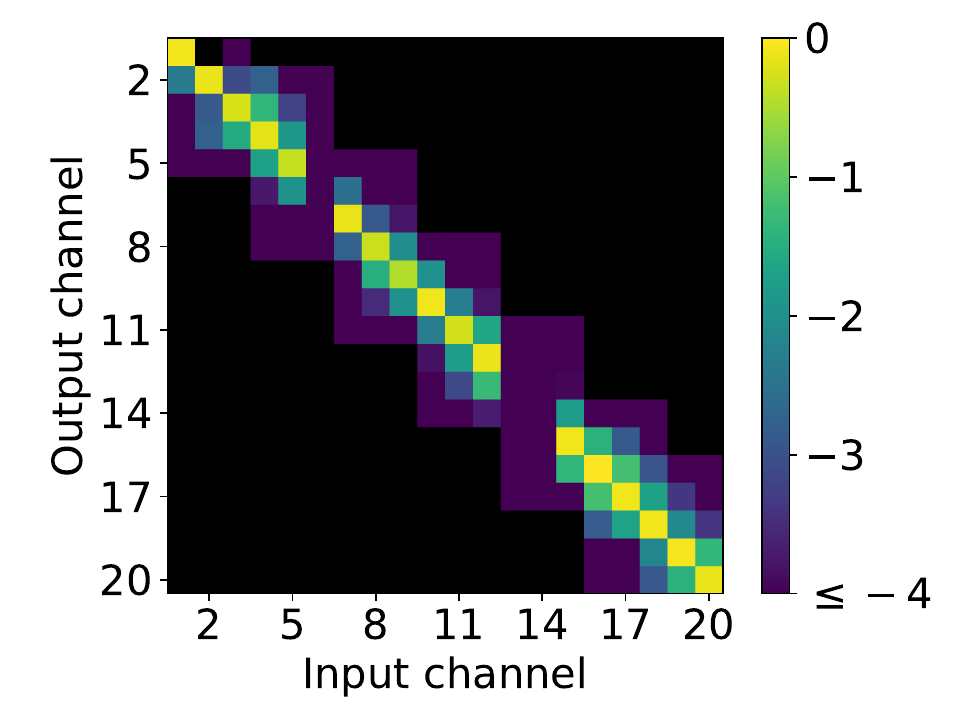}
\caption{Logarithm heatmap of normalised transmission of light through the circuit under identity transformation. Channels 6, 13 and 14 show no measurable transmission due to technical issues of the device. Black cells indicate no value (no measurements were taken for these cells).}
\label{fig:losses}
\end{figure}

\noindent The transmission through the processor under identity transformation was measured using the 925 nm CW laser source and photodiodes. This was done by changing the input through all modes one-by-one and measuring the corresponding output mode as well as the 5 to 7 nearest output modes for each. The normalised output intensity measured is plotted on a log scale in Fig.~\ref{fig:losses} and shows the varying transmission across the different modes as well as the amount of light leaking into neighbouring modes. Channels 6, 13 and 14 show no measurable light throughput due to technical issues of the device. Off diagonal cells plotted in black indicate that no measurements were taken for the corresponding data points. We note that the measured output from channel 1 with channel 2 input has been omitted due to an error in the measurement process. 

Note that the programmable integrated circuit was designed for 940 nm light, and our quantum dot wavelength is 925 nm. We found that performance is better at 940 nm than at 925 nm, as at 940 nm the beamsplitters are closer to the design ideal of 50\% reflectivity, meaning that a given unitary at 940 nm is closer to the ideal unitary than the same circuit at 925 nm. 

\vspace{2mm}
\noindent \textbf{\emph{Calibrating channel efficiencies}}
\vspace{2mm}

\noindent We correct for coupling ($\eta_{\text{c}}$) and single photon detector efficiencies ($\eta_{\text{d}}$) by calculating correction factors $1/\eta_{\text{tot}}$, where $\eta_{\mathrm{tot}}^{-1} \, {=} \, 1/(\eta_{\mathrm{f}} \, \eta_{\mathrm{d}}) \, {\geq} \, 1$, see Table~\ref{tab:channels}.  We investigate the coupling efficiencies with the CW laser and photodiode system.  Light is routed from the reference input (Ch.~1) to all output channels individually, and the resulting voltages are normalised to output Ch.~1, determining $\eta_{\text{c}}$. The overall efficiency $\eta_{\text{tot}}$ is determined in a similar way, with the input a CW weak coherent state and the detection on SNSPDs (Ch.~1--4) and APDs (Ch.~5--8).  The surprising results show significant differences in detector efficiencies $\eta_{\text{d}}$, even within similar architectures.  We attribute this to a combination of varying polarisation sensitivity and efficiency of the detectors, fibre and their interconnection losses, as well as different dark and background noise. The corrected single photon measurements almost perfectly reproduce the CW laser results, indicating the validity of our correction factors. 

\begin{table}[t!]
\resizebox{\columnwidth}{!}{%
\begin{tabular}{|l|l|l|l|l|l|l|l|l|}
\hline
Channel  & 1 (Ref.) & 2    & 3    & 4    & 5    & 6    & 7    & 8    \\ \hline
$\eta_{\text{c}}$    & 1.00                  & 1.02 & 0.91	& 0.93 & 1.02 & 0.95 & 0.76 & 0.82 \\ \hline
$\eta_{\text{d}}$ & 1.00                  & 0.65 & 0.48 & 0.53 & 0.35 & 0.23 & 0.29 & 0.22 \\ \hline
$1/\eta_{\text{tot}}$    & 1.00                  & 1.50 & 2.30 & 2.02 & 2.84 & 4.54 & 4.59 & 5.53 \\ \hline
\end{tabular}%
}
\caption{Efficiencies for couplings ($\eta_{\text{c}}$) and detectors ($\eta_{\text{d}}$), and compensation factors ($1/\eta_{\text{tot}}$) for all eight channels, normalised to channel 1 (Ref.). $\eta_{\text{c}}$ measured with CW laser and photodiodes, $\eta_{\text{tot}}$ with CW weak coherent state and single photon detectors, $\eta_{\text{d}}$ is calculated. \vspace{-6mm}}
\label{tab:channels}
\end{table}

\vspace{2mm}
\noindent \textbf{\emph{Sequential heater optimisation}}
\vspace{2mm}

\begin{figure}[H]
 \centering
\includegraphics[width=\columnwidth]{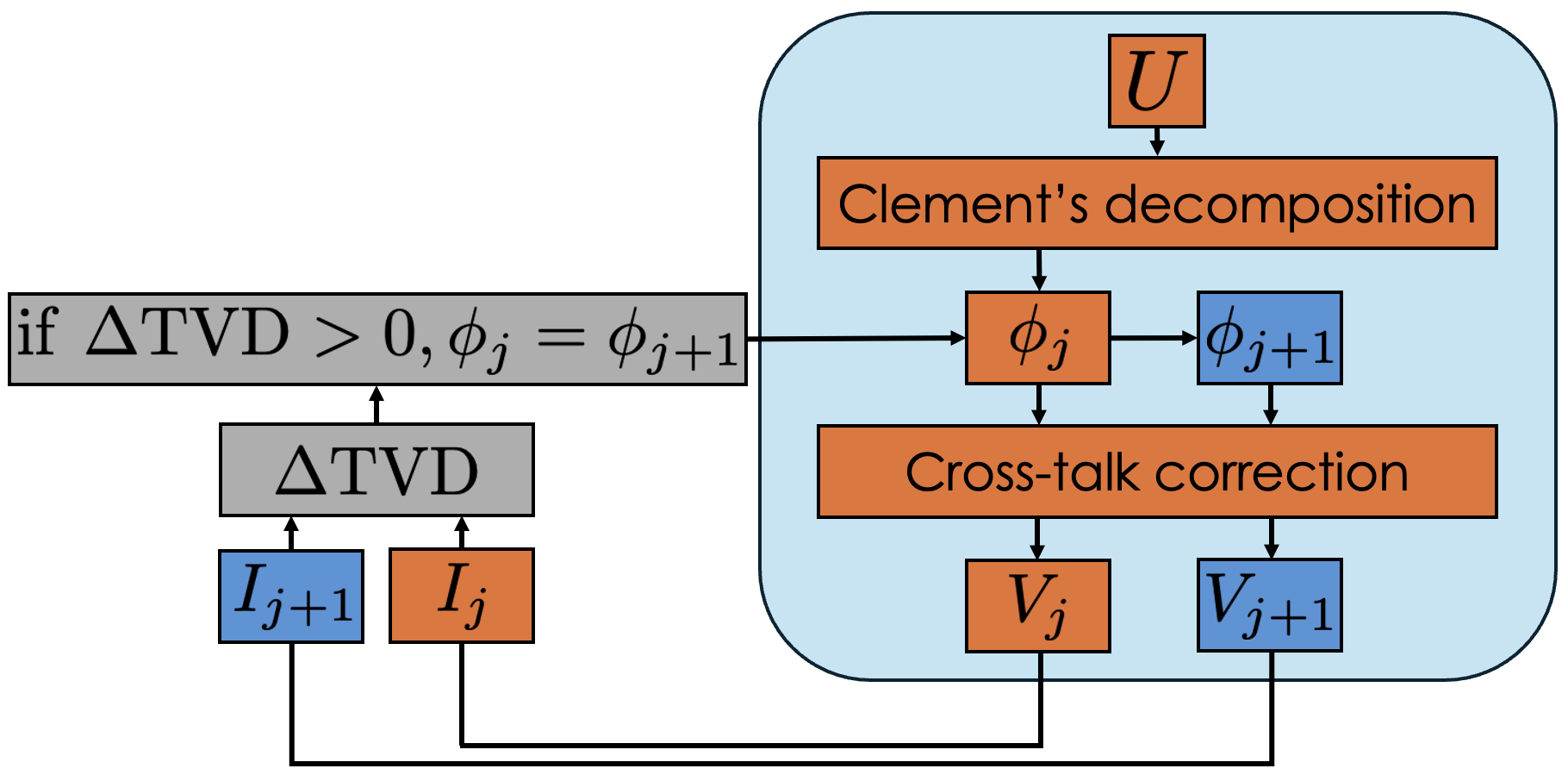}
\caption{Sequential optimisation process flow. The blue box shows the flow for implementing a unitary transformation on the circuit through the device software (orange) and the additional steps for optimisation (blue). Details on the flow can be found in the text. Once the TVD no longer reduces, the process is repeated for all other heaters sequentially.}
\label{fig:processes}
\end{figure}

\noindent The optimisation of the circuit is illustrated in detail in Fig.~\ref{fig:processes}. Starting with a unitary transformation, the initial phases, $\phi_{j}$, are calculated using the processor's Clement's decomposition code within the manufacturer's software. The software then performs cross-talk correction as per the characterisation of the device performed by the manufacturer and maps the these phases onto voltages, $V_{j}$, to set for each phase-shifter. The voltage settings for the initial phases then produce initial output intensity distribution, $I_{j}$. 

The optimisation process targets each individual heater and varies its phase by a small fixed increment from its initial phase, producing phase settings $\phi_{j+1}$. These are then similarly passed through the device software and converted into voltage settings, $V_{j+1}$, producing output intensity distribution, $I_{j+1}$. 

\begin{table*}[t]
\begin{tabular}{|l|l|l|l|l|l|}
\hline
\textbf{Database size (N)}                           & 4               & 5               & 6               & 7               & 8               \\ \hline \hline

\textbf{Success probability} &                 &                 &                 &                 &                 \\ \hline
\phantom{\ \ \ \ \ \ \ Original} a) theory             & 1               & 0.97            & 0.91            & 0.87            & 0.95            \\ 
\ \ \ \ \ \ \ Original b) unoptimised        & $0.94 \pm 0.03$ & $0.86 \pm 0.04$ & $0.78 \pm 0.03$ & $0.75 \pm 0.04$ & $0.79 \pm 0.04$ \\ 
\phantom{\ \ \ \ \ \ \ Original} c) optimised          & N.A. & $0.90 \pm 0.03$  & $0.88 \pm 0.02$ & $0.84 \pm 0.03$ & $0.88 \pm 0.03$ \\ \hline
\phantom{Deterministic} a) theory        & 1               & 1               & 1               & 1               & 1               \\ 

Deterministic b) unoptimised   & $0.94 \pm 0.03$ & $0.91 \pm 0.02$ & $0.89 \pm 0.03$ & $0.86 \pm 0.03$ & $0.82 \pm 0.04$ \\ 

\phantom{Deterministic} c) optimised     & N.A. & $0.92 \pm 0.03$ & $0.90 \pm 0.03$  & $0.90 \pm 0.03$  & $0.88 \pm 0.03$ \\ \hline  \hline

Improvement d) original  & N.A. & 4\% & 10\% & 9\% & 9\% \\ 
\phantom{Improvement} e) deterministic  & N.A. & 1\% & 1\% & 4\% & 6\% \\ \hline
\end{tabular}%
\vspace{-1mm}
\caption{Effect of optimisation on success probabilities: a) theoretical probabilities and b), c) the mean of the measured success probabilities b) before and c) after circuit optimisation. The uncertainty in the mean z is the standard deviation of the measured success probabilities for each $N$. d), e) The difference in success probability due to circuit optimisation for d) original, and e) deterministic, Grover's algorithm. The deterministic algorithm is notably more robust against circuit imperfection than the original algorithm. \vspace{-5mm}}
\label{tab:successprobs}
\end{table*}

The output intensity distributions of the phase settings, $I_{j}$ and $I_{j+1}$, are used to calculate the TVD (Eq.~1 from the main text). If the change in phase of the target heater results in an improvement of the TVD, the initial phases are updated and the process is repeated for the same heater until no further improvement is achieved.  The process is sequentially repeated for all phase-shifters involved in implementing the unitary transformation.

Table~\ref{tab:successprobs} highlights the improvement due to the optimisation routine. We achieved improvements of up to 10\% (6\%) for the original (deterministic) Grover's algorithm. We attribute the lower improvements of the deterministic algorithm to higher initial robustness.

\end{document}